\documentclass[12pt]{article}
\usepackage{amssymb,amsmath}

\input epsf.tex

\title{Mirror symmetry and deformation quantization}
\author {Paul Bressler, Yan Soibelman}

\begin{document}
\maketitle

\newtheorem{thm}{Theorem}
\newtheorem{lmm}{Lemma}
\newtheorem{dfn}{Definition}
\newtheorem{rmk}{Remark}
\newtheorem{prp}{Proposition}
\newtheorem{conj}{Conjecture}
\newtheorem{exa}{Example}
\newtheorem{cor}{Corollary}
\newtheorem{que}{Question}
\newtheorem{ack}{Acknowledgements}
\newcommand{\K}{{\bf k}}
\newcommand{\C}{{\bf C}}
\newcommand{\R}{{\bf R}}
\newcommand{\N}{{\bf N}}
\newcommand{\Z}{{\bf Z}}
\newcommand{\Q}{{\bf Q}}
\newcommand{\G}{\Gamma}
\newcommand{\A}{A_{\infty}}
\newcommand{\ihom}{\underline{\Hom}}
\newcommand{\ra}{\longrightarrow}
\newcommand{\epi}{\twoheadrightarrow}
\newcommand{\mono}{\hookrightarrow}

\newcommand{\epp}{\varepsilon}

\vspace{3mm}

\section{Introduction}

In homological mirror symmetry (see [Ko1]) and in the theory of D-modules  one meets
similar objects. They are pairs $(L,\rho)$ where $L$ is a Lagrangian
manifold, and $\rho$ is a flat bundle on $L$ (local system).
In the framework of homological mirror symmetry such pairs
are objects of the so-called Fukaya category, which
is the principal mathematical structure of the genus zero part of the A-model.
In the framework
of D-modules they are holonomic D-modules (in $C^{\infty}$ category).
It is natural to compare the categories themselves.
This comparison is the main theme of  present paper.

One can object any relationship between mirror symmetry and deformation
quantization. Let us mention some of possible objections.

a) Theory of D-modules (or more generally modules
over the ring of micro-differential operators) ``lives''
on the cotangent bundle $T^{\ast}X$, while the Fukaya
category is defined for any symplectic manifold, hence is more general.

b) Theory of D-modules works well in algebraic or complex
analytic framework, while the Fukaya category (and in general
Floer theory) exists in $C^{\infty}$-category only.

c) In mirror symmetry one considers series in exponentially small (with respect to
the symplectic structure) parameter, while in deformation quantization
the parameter ``is of the size'' of the symplectic structure.

From our point of view these are rather problems  than objections.
In regard to the point a)  this means that one should work
out a version of the theory of microdifferential operators
for arbitary symplectic manifolds. 
 This theory is known as deformation
quantization of symplectic manifolds. 
In regard to the point b) we remark that
finding of a complex analog of the Floer theory is 
an interesting problem. At this time we have only speculations
in this directions. 

The aim of present paper is to summarize our present understanding 
of the topic and make some conjectures. It is a part of an ongoing project.
The details will appear elsewhere.
Our main conjecture can be briefly formulated such as follows:

i) there is a category of holonomic modules over the
 quantized  algebra of smooth functions
on a symplectic manifold;

ii) it is possible to change morphisms in this category by a kind of integral
transformation,
so that it becomes equivalent (at least locally) to the Fukaya
category of the same symplectic manifold.

Let us mention two possible applications of this idea.
First, it can help in constructing of an  algebraic model of the Fukaya category
(analogy: de Rham complex is an algebraic model of Morse complex).
Second, it might help to resolve some difficulties
in the definition of the Fukaya category (non-transversality
of supports, non-existence of identity morphisms, etc.)
In our opinion current situation is not completely satisfactory
despite
the recent progress (cf. for ex. [FOOO]).
Third, one can go beyond Lagrangian submanifolds by considering
coisotropic submanifolds or even non-smooth varieties(cf. [KO]).
Finally, one hopes to achieve a deeper understanding
of the relationship at the level of chiral algebras (see 
[BD],[MSV], [KaV]). Hence the 
subject of this paper is just the first approximation
to the full picture. We hope to discuss it
elsewhere.

The paper is organized as follows. In Section 2 
we briefly recall basics on the deformation theory and $\A$-categories.
The purpose of this section is to fix the language.
Details of the formalism will be explained in [KoSo2].

In Section 3 we recall main facts about the Fukaya category,
which is one of the main structures of homological mirror
symmetry. Section 4 is a reminder on deformation quantization
of symplectic manifolds.
Section 5 is devoted to
the comparison of the Fukaya category with the category of
holonomic modules over the quantized algebra of functions.

Main idea of Section 3 goes back to the pioneering paper [Ko1]. As explained in loc. cit.
homological mirror symmetry is not a statement about
individual categories, but rather  about families of $\A$-categories
over  $\Z$-graded formal schemes.
Geometrically this structure is modelled by a ``family
of non-commutative differential-graded manifolds over a
commutative differential-graded base''. The framework in
which the latter phrase has precise meaning will be explained
in [KoSo2]. The reader can keep in mind the example of the
deformation theory of an associative
algebra. In this case
one has a family of associative algebras parametrized by the formal moduli space
${\cal M}$ of deformations of the given algebra $A$. 
The moduli space ${\cal M}$
is a formal pointed differential-graded manifold (dg-manifold for short)
 (see for ex. [Ko2], [KoSo3]).
Then by definition ${\cal M}$ is a base of a family of associative algebras
$A_{\gamma}, \gamma \in {\cal M}$. Each algebra gives rise to
a ``non-commutative scheme'' $Spec(A_{\gamma})$, hence one has the
desired structure.

{\it Acknowledgements}. We thank to Dima Arinkin and
especially to Maxim Kontsevich for useful
discussions. 
Some of the background material is taken from
the joint book of Kontsevich and second author (see [KoSo2]).
 We thank to Sergei Barannikov 
for comments on the paper.
Y.S. thanks to the Clay Mathematics Institute for
supporting him as a Fellow and to IHES for hospitality
and excellent research conditions. He also thanks to the organizers
of the workshop CATS-1 (Nice, November 2001) where the main 
ideas were reported.

\section{Reminder on $\A$-categories and deformation theory}

In this section we recall some facts about homological algebra of
mirror symmetry. Full details including the
necessary language of non-commutative geometry will appear in [KoSo2].
Some of the material can be found in existing literature,
for example in [Ko2]. We are not going to discuss in detail  motivations 
for all of the definitions and notions below.
Main purpose of this section is to fix the language. 

Let $A$ be a free $\Z$-graded  $k$-module over a unital commutative ring $k$
of characteristic zero (main applications deal with the case when $k$ is a field).

\begin{dfn}
An $\A$-algebra  $A$ over $k$ is given by the following data:

a) A $\Z$-graded free $k$-module $A$.

b) A codifferential $d$ on the
cofree coalgebra
$T(A[1])=\oplus_{n\ge 1}A[1]^{\otimes n}$, where $A[1]$ denotes
the graded free $k$-module such that $A[1]^i=A^{i+1}$ ($A$ with
shifted grading).

(We recall that a codifferential means a coderivation $d$
of the coalgebra satisfying the condition $d^2=0$).

\end{dfn}

Since $d$ is uniquely defined on cogenerators,
it gives rise to ``higher multiplications''
$m_n:A^{\otimes n}\to A, n\ge 1$ of degrees $2-n$ satisfying a
system of quadratic equations which follows from the equation
$d^2=0$.

\begin{dfn}

An $L_{\infty}$-algebra on $A$ 
is given by the following data:

a) A $\Z$-graded vector space $A$.

b) A  codifferential
on the cofree cocommutative coalgebra 
$C(A[1])=\oplus_{n\ge 1}S^n(A[1])$, where $S^n(V)$ denotes the 
$n$th symmetric power in the symmetric monoidal category of $\Z$-graded $k$-modules.

\end{dfn}

The codifferential $d$ defines a sequence of ``higher Lie brackets''
$m_n:A^{\otimes n}\to A, n\ge 1$ of degrees $2-n$ satisfying a system
of quadratic equations which  follows
from the equality $d^2=0$.

It is useful to have in mind geometric picture for both algebraic
structures defined above. We start with  $L_{\infty}$-algebras.

 An $L_{\infty}$-algebra gives rise to a {\it formal pointed
$\Z$-graded manifold $X$}, which carries a vector field $d_X$
of degree $+1$ such that $d_X$ vanishes at the
marked point, and satisfies the condition $[d_X,d_X]=0$. This structure is
called {\it formal pointed differential-graded manifold} in [Ko2],
[KoSo2], [KoSo3] (it was introduced by
A. Schwarz under the name of $Q$-manifold).  Algebra of formal functions
of $X$ is isomorphic to the graded dual to the coalgebra $C(A[1])$.
Thus we can write $X=Spf((C[1])^{\ast})$, where $Spf$ stands for the formal spectrum.

\begin{rmk} a) One should remember that formal $\Z$-graded manifolds have only nilpotent
points.

b) It is useful to interpret maps $m_n$ as Taylor coefficient of the vector field $d_X$
at the marked point.

\end{rmk}

Let us consider two illustrating examples.

\begin{exa} Let $A$ be an associative algebra. Then its 
truncated Hochschild  complex 
$C_{+}^{\bullet}(A,A)[1]=\oplus_{n\ge 1}Hom_k(A^{\otimes n},A)[1]$
carries a structure of differential-graded Lie algebra, hence
defines a formal pointed dg-manifold (see [Ko2]).

\end{exa}

One can interpret the DGLA from the example as a DGLA of coderivations
of the tensor coalgebra cogenerated by $A[1]$. Equivalently, it is
the DGLA of  vector fields  on the formal pointed graded manifold $X$
vanishing at the marked point.
Then the DGLA structure is the natural one on vector fields.

One also has the notion of {\it formal differential-graded manifold},
where the condition of vanishing at the marked point is dropped.
Algebraically this means that we allow the condition $m_0\ne 0$.

\begin{exa} In the previous example we consider the full
Hochschild complex $C^{\bullet}(A,A)[1]=\oplus_{n\ge 0}Hom_k(A^{\otimes n},A)[1]$. It gives rise
to a formal dg-manifold.

\end{exa}

Importance of
$L_{\infty}$-algebras and formal dg-manifolds 
in deformation theory is based on the fact
that they define  deformation functors (see for ex. [Ko1], [KoSo2]).
Let us briefly recall the construction.
If $g=\oplus_{n\ge 0}g^n$ is an $L_{\infty}$-algebra ($g^n$ is
the $n$th graded component) then
one has the deformation functor from commutative nilpotent algebras (possibly graded)  to groupoids.
Namely to a commutative nilpotent ring $R$ one assigns the groupoid
$Def_g(R)$ consisting for $\gamma \in g^1\otimes R$ which satisfy the
Maurer-Cartan equation

$$m_1(\gamma)+m_2(\gamma,\gamma)/2!+...+m_n(\gamma,...,\gamma)/n!+...=0,$$

where $m_n$ are the higher Lie brackets.

Formal deformations of many algebraic and geometric structures
give rise to formal pointed dg-manifolds. Formal dg-manifolds without marked points
arise, for example, when one deforms categories (i.e. when  objects of a category are deformed).
In  Example 2 the corresponding dg-manifold controls deformations of the category
with one object $X$ such that $Hom(X,X)=A$.

One can develop a similar geometric language for $\A$-algebras.
Namely, an $\A$-algebra gives rise to a {\it non-commutative
formal pointed dg-manifold}. It is modelled by the cofree
coalgebra $T(A[1])$ which carries a codifferential $d_X$.
One also has the notion of
{\it non-commutative formal dg-manifold} (no marked point is specified).
In this case one can have a non-zero map $m_0:k\to A$. Geometrically
this structure corresponds to a vector field $d_X$ of degree $+1$
such that 
$[d_X,d_X]=0$, without any condition at the marked point. 
The corresponding algebraic structure is defined in the following way.

\begin{dfn} We say that a codifferential on the coalgebra
$k\oplus T(A[1])$ defines a structure of
generalized $\A$-algebra on $A$.

\end{dfn}

The notion of a small $\A$-category is a natural generalization of the notion of
$\A$-algebra. Traditionally, such a category  is defined by a set of objects $Ob({\cal C})$,
$\Z$-graded free $k$-modules of morphisms $Hom(X,Y)$,
and structures of $\A$-algebras on the
spaces $\oplus_{0\le i,j\le n}Hom(X_i,X_j)$ for any collection
of objects $X_0,...,X_n, n\ge 1$. These structures are 
given by the higher compositions

$$m_n:\otimes_{0\le i\le n}Hom(X_i,X_{i+1})\to Hom(X_0,X_n),$$
which are maps of $\Z$-graded free $k$-modules of degrees $2-n$
satisfying quadratic relations similar to those for $\A$-algebras.
The structures of $\A$-algebras are compatible with inclusions
of collections of objects.

One can think of an $\A$-category ${\cal C}$ as of the large $\A$-algebra
$End(\oplus_{X\in Ob({\cal C})}X)$ (compare with the relation between
additive categories and associative algebras). 
The Hochschild complex and Hochschild
cohomology of an $\A$-category can be defined in terms of this $\A$-algebra.

For any object $X$ the $k$-module $End(X)=Hom(X,X)$
is an $\A$-algebra. Its truncated  Hochschild complex
gives rise to a formal pointed dg-manifold ${\cal M}_X$.
Then the formal dg-manifold ${\cal M}=\sqcup_{X\in Ob({\cal C})} {\cal M}_X$
``controls'' $\A$-deformations of the category ${\cal C}$ with the fixed
set of objects.

Replacing in the above discussion the truncated Hochschild complex
by the full Hochschild complex one obtains a new structure called
{\it generalized $\A$-category}. Generalized $\A$-categories do not
have a fixed set of objects. This is due to the fact that now for
the $\A$-algebra $End(X)$ one can have $m_0\ne 0$.
As before, one can derive the formal dg-manifold ${\cal M}$ 
(now using the generalized $\A$-algebras $End(X)$). For a commutative
nilpotent $k$-algebra  $R$ one can consider
$k$-points ${\cal M}(R)$. 
If ${\cal M}_0\subset {\cal M}$ is the
subset of zeros of the odd vector field $d_{\cal M}$ then one can speak
about objects of some $\A$-category.  The objects are parametrized by ${\cal M}_0$.
We omit here the description which can be given
in terms of $R$-points of  ${\cal M}_0$ and ${\cal M}$.

We will keep the name of generalized $\A$-category
for a slightly more general structure. The point
is that we allow the higher compositions $m_n$  be defined not for all
collections of objects, but only for some of them
(transversal collections). The Fukaya category discussed
in the next section will be this kind of generalized $\A$-category.

We  summarize without details the data defining a generalized $\A$-category 
in the following way:

a) We are a given a formal dg-manifold  $Ob({\cal C})={\cal M}$.

b) For any $n\ge 1$ we are given a formal dg-submanifold 
${\cal M}^n_{tr}\subset {\cal M}^n$ called the space of transversal $n$-families.
It is assumed that ${\cal M}^1_{tr}\subset {\cal M}$ (i.e.
every object is transversal to itself).

c) For a point  $(X,Y)\in {\cal M}^2_{tr}$ we are given a $\Z$-graded
free $k$-module $Hom_{\cal C}(X,Y)$ called a space of morphisms between
$X$ and $Y$. These $k$-modules are organized in a formal dg-bundle
$Hom^{\cal C}\to {\cal M}^2_{tr}$.

d) For $n\ge 1$ and any  $(X_0,...,X_n)\in {\cal M}^n_{tr} $ we are given
a higher composition map $m_n^{(X_0,...,X_n)}$ which gives rise to a morphism
of the obvious pullbacks of $Hom^{\cal C}$  to ${\cal M}^n_{tr}$.
The  composition maps
give rise to a structure of generalized $\A$-algebra $A(X_0,...,X_n)$,
or, equivalently to a non-commutative formal 
dg-manifold ${\cal M}(X_0,...,X_n)$.

e) Let  ${\cal M}({\cal C})$ be the inductive limit 
of ${\cal M}(X_0,...,X_n)$ taken over increasing collections
of transversal objects. This is a non-commutative formal dg-ind-scheme
(it can be properly defined as an inductive limit in the appropriate
category).

f) Finally  ${\cal M}({\cal C})$ is a non-commutative formal dg-manifold
over the commutative scheme $Spec(k)$. 

The  structure defined in a)-e)  is called generalized
$\A$-category. It gives rise to the usual $k$-linear
$\A$-category, which we will
call associated to ${\cal C}$. The latter is defined by
a non-commutative formal dg-ind-subscheme 
${\cal M}({\cal C})_0\subset {\cal M}({\cal C})$ of zeros
of the odd vector field $d_{{\cal M}({\cal C})}$.

\begin{rmk} a) One can define the Hochschild complex of
a generalized $\A$-category. It gives rise to a 
formal dg-manifold. 
The corresponding deformation functor describes the formal deformation theory
of the generalized  $\A$-category in the class of generalized $\A$-categories.
If the category has one object $X$ we have 
the full Hochschild complex of the corresponding generalized
$\A$-algebra $End(X)$.

b) We do not discuss  the delicate problem of unital $\A$-algebras
(more generally, $\A$-categories with identity morphisms). It is an interesting question because
in practice  identity morphisms may be defined  up to a homotopy only.
All this can be formulated in the language of non-commutative geometry.

\end{rmk}

There is a notion of $\A$-functor between two generalized $\A$-categories.
$\A$-functors form themselves a generalized $\A$-category.
Using $\A$-functors one defines the notion of equivalence
of $\A$-categories (see [KoSo2]).

Let us illustrate the  notion of equivalence in the case when
both categories have only one object.
Then we are dealing with generalized $\A$-algebras, say $A$ and $B$.
Assume in addition that $m^A_0=0$ and $m_0^B=0$ (i.e. we have
ordinary $\A$-algebras).
If $A$ is equivalent to $B$ then the complexes
$(A,m_1^A)$ and $(B,m_1^B)$ are quasi-isomorphic.
Geometrically this means that the tangent 
spaces at marked points 
of equivalent non-commutative formal pointed dg-manifolds
are quasi-isomorphic. The  converse is also true (this is the $\A$-version of the
inverse function theorem). A generalized $\A$-algebra with $m_0\ne 0$
is equivalent to one which has $m_0\ne 0$, and $m_{n\ge 1}=0$
(cf.: vector field which is non-trivial at a point
is locally equivalent to a constant one).
The latter observation explains 
why generalized $\A$-categories should be studied
in families rather than individually. Indeed generalized $\A$-categories
with $m_0\ne 0$ are trivial in the sense that all higher compositions
can be killed by an appropriate equivalence functor.

 Similarly to the case of formal pointed dg-manifolds
there is a theory of minimal models of non-commutative formal 
pointed dg-manifolds. For such a theory one needs to assume that the ground ring $k$ is a field
of characteristic zero (there is more complicated  theory for non-pointed
dg-manifolds).

 If two $\A$-categories are equivalent then the
formal pointed dg-manifolds of their deformations are quasi-isomorphic
(i.e. tangent complexes at the marked points are quasi-isomorphic).

Finally, there is a theory of generalized $\A$-categories
over a formal dg-base. In the above discussion we discussed
 the case when the base was an ordinary scheme $Spec(k)$. We can also assume that
the base is a formal scheme.  As we will see in the next section the latter case is important
in symplectic geometry.

\section{Fukaya category}

\subsection{Fukaya category and non-commutative geometry}

The Fukaya category $F(X)$ of a smooth symplectic manifold $X$
is a generalized $\A$-category over a base. It can be constructed
as an $\A$-deformation of the following trivial $\C$-linear category $F_0(X)$.
Objects of $F_0(X)$ are pairs $(L,\rho)$, where $L$ is a Lagrangian
submanifold of $X$ and $\rho$ is a local system on $L$.
Transversal collections of objects correspond to transversal
collections of Lagrangian submanifolds (in fact we need more sofisticated
transversality condition, see [KoSo1]).
We set 
$Hom_{F_0(X)}((L_0,\rho_0),(L_1,\rho_1))=\oplus_{x\in L_0\cap L_1}
Hom(\rho_{0x},\rho_{1x})$ and
$Hom_{F_0(X)}((L,\rho),(L,\rho))=\Omega^{\bullet}(L,End(\rho))$.
All compositions $m_n, n\ge 1$ are trivial for collections
of different objects. Otherwise $m_{n\ge 3}=0$, and $m_2$  is the natural product
on the differential-graded algebra  $\Omega^{\cdot}(L,End(\rho))$.

To a pair $(L,\rho)$ one can associate a generalized $\A$-algebra
$A(L,\rho)$. Let assume for simplicity that $\rho$ is a trivial
rank one local system. Then the corresponding generalized $\A$-algebra $A(L)$
is generated by geometric cycles in $L$. Higher multiplications
$m_n(C_1,...,C_n)$ between generic cycles $C_i$ are
given by a kind of quantum cohomology construction. Namely,
one counts with some weight 
pseudo-holomorphic maps $f:(D^2,\partial D^2)\to (X,L)$ 
with marked points $x_1,...,x_n\in \partial D^2$.
Here $D^2\subset \C$ is the standard disc. It is required 
that the point $x_i$ is mapped to $C_i$. The weight is
$exp(-{1\over{\epsilon}}\int_{D^2}f^{\ast}(\omega))$, where
$\omega$ is the symplectic form on $X$, and $\epsilon$ is a parameter.
The idea of this construction of $A(L)$ was suggested by Kontsevich.
Difficult analytic details have been worked out in [FOOO].
In what follows we will assume the conditions on the data imposed in  [FOOO]. 
The resulting generalized $\A$-algebra $A(L)$ is defined
over the valuation ring 
$\C_{\epsilon}^{\ge 0}=\{f=\sum_{i\ge 1}a_ie^{-\lambda_i/\epsilon}\}$,
where $a_i\in \C,a_1\ne 0$,
and $\lambda_i\ge 0$ is a monotonically increasing sequence of real
numbers such that $lim_{i\to +\infty}\lambda_i=+\infty$.
The valuation map is given by $v(f)=\lambda_1$.
The corresponding valuation field $\C_{\epsilon}$ consists of series $f$
as above,
with $\lambda_i\in \R$. It is useful to notice
that $\C_{\epsilon}^{\ge 0}$ contains the maximal ideal 
$\C_{\epsilon}^{>0}$ consisting
of series with all $\lambda_i>0$.
One observes that the composition $m_0=0$ modulo $\C_{\epsilon}^{>0}$, but in general
$m_0\ne 0$.

It can be proved that the disjoint union of non-commutative dg-manifolds
associated with generalized $\A$-algebras $A(L,\rho)$ gives rise
to a generalized $\A$-category $F_{\epsilon}(X)$ over the 
formal spectrum of the ring
$\C_{\epsilon}^{\ge 0}$. Its reduction modulo the ideal
$\C_{\epsilon}^{>0}$ is equivalent to $F_0(X)$.
If the symplectic form $\omega$ satisfies certain rationality
conditions then one can introduce a new
parameter $q=exp(-{1\over{\epsilon}})$,
thus replacing the ring $\C_{\epsilon}^{\ge 0}$ by the ring of formal series
$\C[[q]]$. Then the maximal ideal is just $q\C[[q]]$, and the  field
$\C_{\epsilon}$ coincides with the field of Laurent series $\C((q))$.

The Fukaya category $F(X)$ is defined as an $\A$-category obtained
from $F_{\epsilon}(X)$ by restriction to zeros of the odd vector field.
In particular, for an object $A$ of $F(X)$ the composition $m_0$ vanishes:
$m_0^A=0$. The condition $m_0^A=0$ defines a 
``subvariety'' of the non-commutative
moduli space ${\cal M}^{NC}_{Ob(F_{\epsilon}(X))}$. Objects of the Fukaya
category exist only along this ``subvariety''. This geometric picture
explains why
it is too naive to work with the Fukaya category for fixed $\epsilon$,
even if one can prove convergence of the series defining $m_n$
(the latter is still an open problem).

Let us assume for simplicity
the above-mentioned rationality conditions of $\omega$. Then the Fukaya
category is defined over
the formal spectrum $Spf(\C[[q]])$. In fact one can extend the definition
so that the base will be 
$Spf(\C[[q]]\otimes(\otimes_{i\ne 2} \C[t_{i,\mu}]))$, 
where $q$ has degree zero, and
$t_{i,\mu}$ are parameters of degrees $2-i$ corresponding to some basis 
in the graded vector space of cohomology
$H^{i}(X,\C)$. We introduce new parameters $z,t$ of degree zero
by setting $z=qe^t$.
Then, inverting $z$ we obtain a family of $\A$-categories over 
the field $\C((z))$
parametrized by $Spf(\C[[t]]\otimes (\otimes_{i\ne 2}\C[t_{i,\mu}]))$. 
This family should be thought of as formal deformation of a certain $\A$-category
over $\C((z))$. The tangent space to the moduli space of the formal
deformations of this category is isomorphic to the cohomology 
$H^{\bullet}(X,\C((z)))$. 
The latter cohomology group is isomorphic
to $\oplus_{i\ge 0}Ext^i(Id,Id)$, where $Id$ is the identity functor,
and the extensions are taken in the properly
defined $\A$-category of endofunctors.
This description is useful for the purposes
of quantum cohomology (Yoneda product on functors gives rise
to the quantum product on the cohomology group).

\subsection{Conventional approach to the Fukaya category}

Below we briefly recall the ``naive'' definition of the Fukaya category,
when the composition $m_0$ is ignored.
It is useful in some questions, for example in mirror
symmetry for abelian varieties (see [KoSo1]). As we will discuss below, this
``naive'' Fukaya category is related to deformation quantization.

Objects of $F^{naive}(X)$ are pairs $(L,\rho)$, where
$L\subset X$ is a Lagrangian submanifold and
$\rho$ is a local system on $L$. Morphisms
between $(L_0,\rho_0)$ and $(L_1,\rho_1)$ are defined
only if $L_0$ and $L_1$ intersect transversally.
In this case $Hom((L_0,\rho_0),(L_1,\rho_1))=
\oplus_{x\in L_0\cap L_1}Hom(\rho_{0x},\rho_{1x})\otimes {\C_{\epsilon}}$.
The space of morphisms is $\Z$-graded by means of Maslov index.
Thus we are dealing with graded Lagrangian
manifolds (cf. [Se1]).
There are higher compositions $m_n,n\ge 1$, which are
linear maps of degrees $2-n$:

$$m_n:\otimes_{0\le i\le n}Hom((L_i,\rho_{i}),(L_{i+1},\rho_{i+1}))\to
Hom((L_0,\rho_0),(L_n,\rho_n)).$$
They are defined by means of the Floer-type construction
associated with the ``transversal'' collection of 
Lagrangian submanifolds $L_i,0\le i\le n$. It is usually said that
the maps $m_n$ give rise to an $\A$-structure on $F^{naive}(X)$.
We refer the reader to [KoSo1] about the details of this definition
and to [FOOO] about definitions of the related moduli spaces.

There are several problems with the naive definition
of the Fukaya category. One of the most essential is the presence
of pseudo-holomorphic discs with the boundary mapped to a Lagrangian
submanifold. This amounts to non-trivial maps
$m_0:\C_{\epsilon}\to Hom(X,X)$. As a result, the axioms of $\A$-category
are not satisfied, and one has to work with generalized $\A$-categories.
It was explained in the preceeding sections (for all details see  [KoSo2]) how a
consistent theory of such can be developed
in the framework of non-commutative formal geometry.
One can still work with $F^{naive}(X)$ with understanding that it is
only a part of the ``true'' Fukaya category $F(X)$.

\section{Reminder on deformation quantization}

We recall that a symplectic 
2n-dimensional manifold $(X,\omega)$ gives rise to 
an abelian category ${\cal C}(X)$ of modules over a non-commutative
algebra $A(X)$ (deformation quantization
of the algebra $C^{\infty}(X)$ of smooth functions on $X$).
Such a deformation quantization is non-unique. We will use the one 
which has the characteristic class $[\omega]/t$ (see [De]).
The algebra $A(X)$ is a topological algebra over 
the ring of formal series $\C[[t]]$. As $\C[[t]]$-module it is isomorphic
to the algebra of formal series $C^{\infty}(X)[[t]]$.
The algebra $A(X)$ 
consists of global sections of a sheaf of non-commutative algebras
$A_X$, such that locally $A_X$ is 
isomorphic to the sheaf of $t$-pseudo-differential
operators on $\R^n$ (the latter are locally series 
 $P=\sum_{|I|\ge 0}a_I(x)(t\partial_x)^I$). The Poisson structure
 induced on $C^{{\infty}}_{X}\simeq A_{X}/tA_{X}$ coincides with the 
 one given by the symplectic form.

One has a category of $A(X)$-modules $M$ 
 such that $M$ is t-adically complete,
flat as $\C[[t]]$-module, and $M/tM$ is
the space of sections of a sheaf of modules over the sheaf 
of smooth functions $C^{\infty}_X$. 
The category of $A_X$-modules  will be denoted by
${\cal C}(X)$.
 Morphisms are defined
as $\C[[t]]$-linear homomorphisms of topological modules.
We will keep the same notation for  the related category 
 defined
over the field $\C((t))$. It is obtained from
${\cal C}(X)$ respectively ${\cal C}_X$ by adding $t^{-1}$, so that modules $V$ and $tV$ become
equivalent.

Let $hol(X)$ be a full subcategory of ${\cal C}(X)$
which consists of modules $M$ such that the support $Supp(M/tM)$ is
a Lagrangian submanifold.

We will call objects of $hol(X)$ 
{\it holonomic}.
The Lagrangian support $Supp(M)$ of a holonomic module will be sometimes called its
{\it characteristic
variety} of $M$ and denoted by $Ch(M)$. The 
category $hol(X)$  contains
objects  $V_{(L,\rho)}$ which correspond to pairs
$(L,\rho)$ where $L\subset X$ is a Lagrangian submanifold and
$\rho$ is a local system on $L$. In what follows only such objects
will be considered. 

\begin{rmk}
 In [KaMa] the authors constructed (for every Lagrangian submanifold $L$ satisfying
some topological conditions)
an $A_X$-module $V_L$ such that $Ch(V_L)=L$. 
One can easily generalize their construction including local systems on $L$.
We will call the corresponding objects
Karasev-Maslov modules.

\end{rmk}

We will  need symplectic manifolds $X^n, n\ge 2$. The corresponding symplectic forms
are given by $(\omega,-\omega,-\omega,...,-\omega)$.

The identity functor $Id_{A_X-mod}$ is represented by the  $A_{X\times X}$-module
$K_{\Delta}$  supported on the diagonal $\Delta\subset X\times X$.
It can be  identified with the sheaf  $A_X$.
Deformations of $A_X-mod$ as an $\A$-category has the tangent complex
quasi-isomorphic (after a shift) to the tangent
complex to the  deformations of the identity functor $Id_{A_X-mod}$.
The latter deformations are described by the deformations
of $K_{\Delta}$ as an object of $A_{X\times X}-mod$.  The tangent space at  $K_{\Delta}$
 to the moduli space of its 
deformations is isomorphic to $\oplus_{i\ge 0}Ext_{A_{X\times X}-mod}^i(K_{\Delta},K_{\Delta})$.
After changing scalars to $\C((t))$ the latter sum can be identified with $H^{\bullet}(X,\C((t)))$.
We can restrict the deformation functor to the subcategory $hol(X)$. It is not difficult
to see that the support of a holonomic module remains Lagrangian. 
These observations lead to the following result.

\begin{prp} 
a) The Hochschild cohomology of the category 
$A_X-mod$ is isomorphic to  $H^{\bullet}(X,\C((t)))$.
(Here we consider $A_X-mod$ as an $\A$-category with $m_{n\ne 2}=0$ and $m_2$ given
by the usual composition of morphisms).

b) The tangent space to the deformations of an object $(L,\rho)\in hol(X)$ is
isomorphic to  $H^{\bullet}(L,End(\rho))\otimes \C((t))$.

\end{prp}

We also mention the following proposition (see [So1]). It should be compared
with the definition of $Hom's$ in the Fukaya category.

\begin{prp} Let $V_{(L,\rho)}$ denotes the object of $hol_X$
correspodning to the pair $(L,\rho)$. If $L_0$ is transversal
to $L_1$ then 

a) $Ext^i(V_{(L_0,\rho_0)},V_{(L_1,\rho_1)})$ is trivial
if $i\ne n$, where $n=1/2\,dim\,X$;

b) $Ext^n(V_{(L_0,\rho_0)},V_{(L_1,\rho_1)})\simeq
\oplus_{x\in L_0\cap L_1}Hom(\rho_{0x},\rho_{1x})\otimes \C((t))$;

c) the algebra $Ext^{\bullet}(V_{(L,\rho)},V_{(L,\rho)})$
is isomorphic to the cohomology $H^{\bullet}(L,End(\rho))\otimes \C((t))$
(cf. part b) of the previous Proposition).

\end{prp}

 Let $D^{b}_{\infty}(hol({X}))$ be the $\A$-category
 associated with the category $hol({X})$. 
It is in fact a dg-category. In order to construct it one
choses  injective resolutions of $A_X$-modules  $I_M$ and $I_N$  of two 
holonomic modules $M$ and $N$. Then one defines
 $Hom_{D^{b}_{\infty}(hol({X}))}(M,N)=Hom^{\bullet}(I_M,I_N)$.
In this way one obtains an $\A$-model for the derived category
of the category of holonomic modules.

The discussion above shows certain similarity between 
$D^{b}_{\infty}(hol(X))$ and $F^{naive}(X)$. At the same time their
deformation theories 
induce different products on the cohomology of $X$. In case of the 
Fukaya category it is the quantum product, while in case of $hol({X})$ it is
the usual cup product. The reader  will also notice that Maslov index is not
visible in the case of $hol(X)$.

On the other hand we will explain in the next section that:

a) The algebras $End_{D^{b}_{\infty}(hol(X))}((L,\rho))$ 
and  $End_{F^{naive}(X)}((L,\rho))$ are $\A$-equivalent.

b) If $L$ and $L^{\prime}$ are Hamiltonian isotopic, then we can
``twist'' the space 
$Hom_{D^{b}_{\infty}(hol(X))}((L,\rho),(L^{\prime},\rho^{\prime}))$ in such a way that it becomes
quasi-isomorphic to the corresponding complex of morphisms in $F^{naive}(X)$.

\section{Comparison of the categories}
 
Let $(L,\rho)$ be a pair as before, i.e.
$L$ is a Lagrangian submanifold of $X$
and $\rho$ is a local system on $L$. Let us denote by $E_{(L,\rho)}$
the correspodning object of $F(X)$, and by $V_{(L,\rho)}$ the corresponding
object of $hol(X)$. 
We assume that  $m_0=0$ in the
$\A$-algebra $A(L,\rho)$. This means that in fact we are dealing
with the category $F^{naive}(X)$. We also assume the conditions
imposed on $L$ in [FOOO]. This allows us to make necessary
choices without further explanations. In particular $L$ is relatively spin
in the sense of [FOOO], so the moduli spaces of pseudo-holomorphic
discs are orientable. Taking the Hochschild complex of $A(L,\rho)$
we can construct the formal pointed dg-manifold ${\cal M}_{E_{(L,\rho)}}$
of deformations of $E_{(L,\rho)}$.
Similarly we can start with the Lie
 algebra $Hom_{D^{b}_{\infty}(hol(X))}(V_{(L,\rho)},V_{(L,\rho)})$
and construct the formal pointed dg-manifold
${\cal M}_{V_{(L,\rho)}}$ of deformations of $V_{(L,\rho)}$.

Let us imagine that both $F^{naive}(X)$ and $D^{b}_{\infty}(hol(X))$ are ``sheaves of 
$\A$-categories'' on the ``moduli space of objects''.
 Let us also imagine that there is
a well-defined moduli space $Lagr_X$ of Lagrangian submanifolds
of $X$. Then we should have a natural projection
$\pi:{\cal M}_X\to Lagr_X$. If $L$ is a Lagrangian submanifold,
and $[L]\in Lagr_X$ the corresponding point of the moduli space then the fiber
$\pi^{-1}([L])$ consists of local systems supported on $L$
(or on any Lagrangian submanifold representing the same equivalence class
in the moduli space). We would like to compare $F^{naive}(X)$ and $hol(X)$
in a ``small neighborhood of $[(L,\rho)]$''. From the categorical 
point of view we have two $\A$-categories ${\cal A}$ and ${\cal B}$ with the same ``space'' of 
objects, and such that for any object $X$ the $\A$-algebras $End_{\cal A}(X)$
and  $End_{\cal B}(X)$ are equivalent.
We would like to find a functor $\Phi:{\cal A}\to {\cal A}$ such that
changing morphisms in ${\cal A}$ to $Hom_{\cal A}^{new}(X,Y)=
Hom_{\cal A}(X,\Phi(Y))$ one gets a new $\A$-category equivalent to ${\cal B}$.

\subsection{Main conjecture}

We will denote by $hol(L)$ the full subcategory
of $hol(X)$ consisting of holonomic modules with the given
support $L$. For simplicity we will assume the rationality condition
imposed on the symplectic form $\omega$. Hence the Fukaya category
is defined over $\C((q))$ (this is not a serious restriction because one can consider
deformation quantization over any pro-nilpotent algebra, in particular $\C_{\varepsilon}$).

Our main idea can be explained such as follows. Having in mind the intuitive picture of the previous
subsection we consider both $F^{naive}(X)$ and $D^b_{\infty}(hol(X))$ in a small
neighborhood of a given $[L]\in Lagr_X$. For a pair $L_1,L_2$ sufficiently close to
$L$ we would like to find a functor 
$\Phi_{L_1,L_2}:D^b_{\infty}(hol({L_2}))\to D^b_{\infty}(hol({L_1}))$, such
that $Hom(\rho_1,\Phi_{L_1,L_2}(\rho_2))$ (morphism as objects of
$D^b_{\infty}(hol({L_1})$)) is quasi-isomorphic to $Hom_{F^{naive}(X)}((L_1,\rho_1),(L_2,\rho_2))$.
Such a functor should be represented by a bimodule. Let us describe all this more precisely.

Let $M_{i}=V_{(L_i,\rho_i)}\in hol({L_i}), i=1,2$.
We expect there exists a Lagrangian submanifold $\Lambda_{12}=\Lambda(L_1,L_2)\subset
X\times X$ and $K(L_1,L_2)=K_{\Lambda_{12}}\in hol({X\times X})$ such that:

1) If $L_1$ and $L_2$ have a non-empty intersection then
 $\Lambda_{12}\circ L_1=L_2$. Here $\Lambda\circ L=
\pi_2(\pi^{-1}_1(L)\cap \Lambda)$, where $\pi_i:X\times X\to X, i=1,2$ are
the natural projections. In particular we assume that the restrictions of
$\pi_i, i=1,2$ to $\Lambda$ are coverings.

2) $Hom_{D^b_{\infty}({hol(X}))}(M_{1},K(L_1,L_2)\circ M_{2})\simeq
Hom_{F^{naive}(X)}((L_1,\rho_{1}),(L_2,\rho_{2}))$, where 
 $\simeq$ means  quasi-isomorphism
of complexes, and we consider both categories over the field $\C((q))$
(i.e. $q=t$ in the case of $hol(X)$).
The composition $\circ$ for modules
is given by the formula
$K\circ M=\pi_{2\ast}(K\otimes \pi_1^{\ast}(M))$.
 
We denote by $Hom^{new}(M_{1},M_{2})$ the left hand side of 2).

3) For a generic sequence of Lagrangian submanifolds $L_1,L_2,...,L_n, n\ge 2$
and holonomic modules $M_{1},...,M_{n}$ such that $M_{i}=V_{(L_i,\rho_i)}$
for all $i$, we expect to have an isomorphism of ${A}_{X^n}$-modules

$$K(L_1,L_2)\circ K(L_2,L_3)\circ...\circ K(L_{n-1},L_n)
\to K(L_1,L_n).$$

Such an isomorphism defines a linear map

$$m_n^{new}:\otimes_{1\le i\le n-1}Hom^{new}(M_{i},M_{{i+1}})\to
Hom^{new}(M_{1},M_{n}).$$

We expect the above data satisfy the following

\begin{conj} (i)  Higher compositions $m_n^{new}, n\ge 1$ give rise
to a structure of an  $\A$-category on $D^b_{\infty}(hol(X))$.

(ii) This category is $\A$-equivalent to $F^{naive}(X,\omega)$ (with $q=t$).

\end{conj}

\subsection {The conjecture in the case of cotangent bundle}

Let us fix a Lagrangian submanifold $L\subset X$, and consider
only those $L^{\prime}$ which are ``very close'' to $L$.
More precisely we assume that they are not only close
to $L$ but also Hamiltonian isotopic to $L$.
We want to  ``restrict'' $F^{naive}(X)$ to this ``neighborhood of $L$''.
This means that we consider an $\A$-subcategory with the
objects taken from the above-mentioned subset, and morphisms
same as in $F^{naive}(X)$. We do the same thing with $hol(X)$ and 
$D^b_{\infty}(hol(X))$. We would like to compare these categories
in the case when $X=T^{\ast}Y$ is the cotangent bundle
with the standard symplectic structure (notice that a neighborhood of a
Lagrangian submanifold $L$ can be identified by a symplectomorphism
with a neighborhood of the zero section in $T^{\ast}L$).
We are going to consider Lagrangian
submanifolds of the type $L_i=\{(x, df_i(x))|x\in Y\}$.
Let $\rho_i$ be local systems on $L_i$.

For a pair of such Lagrangian submanifolds we have a symplectomorphism
$\phi:X\to X$ such that $(x,\xi)\mapsto (x, \xi+df_2(x)-df_1(x))$. 
Clearly it maps isomorphically
$L_1$ into $L_2$ .

Let us define $\Lambda=\Lambda_{12}$ as $graph(\phi)\subset X\times X$.
The corresponding bimodule $K_{\Lambda}$ is 
the quotient of $A_X\boxtimes A_X^{op}$ by the 
left ideal generated
by the relation $a\otimes 1=1\otimes e^{{1\over t}ad(f_2-f_1)}(a)$, $a\in A_X$.
Here $ad(a)(b)=ab-ba$ (clearly $A_X$ contains $C_Y^{\infty}$ as a subalgebra,
so the ideal is well-defined).

Notice that 

$exp{1\over t}(ad(f_2-f_1))exp{1\over t}(ad(f_3-f_2))...
exp{1\over t}(ad(f_n-f_{n-1}))=exp{1\over t}(ad(f_n-f_1))$. 
Hence we have an
isomorphism $K(L_1,L_2)\circ K(L_2,L_3)\circ...\circ K(L_{n-1},L_n)
\to K(L_1,L_n)$.

In order to check the conjecture we may assume that $f_1=0$. Then we observe that
$$Hom_{D^b_{\infty}(hol(X))}(\rho_1,K(L_1,L_2)\circ \rho_2)=
\Omega^{\bullet}(Y,\rho^{\ast}_1\otimes \rho_2)$$
 where the RHS
is the complex of de Rham forms with values in the local system. Let $\nabla_i, i=1,2$
denotes the flat connection on $\rho_i, i=1,2$. Then the
differential is given by $\nabla_1^{\ast}\otimes 1+1\otimes \nabla_2+
df_2\wedge (\cdot)$. The latter complex is equivalent to the standard
de Rham complex (without $df_2$) if one twists the sections by
$exp({1\over t}f_2)$, i.e. $s\mapsto s\,exp({1\over t}f_2)$. Then according to [KoSo1], Section 4
the resulting complex is quasi-isomorphic to 
$Hom_{F^{naive}(X)}((L_1,\rho_1),(L_2,\rho_2))$. 
We remark that in the notation of [KoSo1] one has $q=exp(-{1\over \varepsilon})$.

\begin{rmk} The case of general symplectic manifold does not follow
automatically from the results of this section. Indeed, in order
to define morphisms in the Fukaya category for
Lagrangian submanifolds in a small neighborhood of a given
$L$  one has to consider pseudo-holomorphic
discs which do not belong entirely to the neighborhood (we are restricted by the boundary
conditions only). Nevertheless, if our main conjecture is true, one can find
a family of kernels $K(L_1,L_2)$ which takes care about such discs.

\end{rmk}

\subsection{Complex structure on the moduli space of holonomic modules}

We observe that the tangent space to the moduli space of
deformations of a module $M\in hol(X), supp(M)=L$ is isomorphic 
to $Hom_{D^b_{\infty}(hol(X))}(M,M)$ (derived deformations of $Hom_{hol(X)}(Id_X,Id_X))$.
There is a natural embedding $Hom_{D^b_{\infty}(hol(L))}(M,M)\to Hom_{D^b_{\infty}(hol(X))}(M,M)$,
corresponding to the deformations with the fixed support $L$.
On the other hand there is a natural projection
$Hom_{D^b_{\infty}(hol(X))}(M,M) \to \Omega^{\bullet}(L)$, where  $\Omega^{\bullet}(L)$
denotes the de Rham complex of $L$. Indeed,
a deformation of the module $M$
induce the deformation of the support of $M$. The latter
are controlled by differential forms on the support.
We can perform computations in the derived categories.
Then we have an exact sequence of the tangent spaces
to the formal moduli spaces of deformations:

$$ Ext^{\bullet}_{hol(L)}(M,M)\to Ext^{\bullet}_{hol(X)}(M,M)\to H^{\bullet}_{DR}(L).$$

Suppose that $M$ is a simple module. Then
$$RHom_{hol(L)}(M,M)\simeq
R\Gamma(L,R\underline{Hom}(M,M))\simeq R\Gamma(L,{\C}_L)
\simeq \Omega^{\bullet}(L).$$

Taking first cohomology (this corresponds to ``classical" tangent space)
we obtain in this case an exact sequence

$$H^1_{DR}(L)\to Ext^1_{hol(X)}(M,M)\to H^1_{DR}(L).$$

This means that the tangent space to the ``classical" deformations
of $M$ inside of $hol(X)$ is twice as big as the tangent space to the
``classical" deformations of $L$ inside of $Lagr_X$.
If $X$ is a Calabi-Yau manifold then one hopes to obtain a complex
structure on the moduli space of formal deformations of a simple
holonomic module $M$. We expect it is isomorphic to a subvariety in the dual 
Calabi-Yau manifold.
In order to describe the complex structure explicitely
we need to identify the tangent
space $T_L(Lagr_X)$ with the tangent space $T_M(hol(L))$.
It is sufficient to describe the lifting of paths from $Lagr_X$ to
$hol(L)$. Given a path $L(t)\subset Lagr_X$ such that $L(0)=L$,
we define a path $M(t)\subset hol(L), M(0)=M$ in the following way:
$M(t)=M\otimes \rho(t)$. Here $\rho(t)$ is the restriction to $L(t)$
of the  unitary bundle over $X$
with a connection $\nabla$ such that $curv(\nabla)=\omega_X$
(it is often called the pre-quantum line bundle).
Restriction of $\nabla$ to $L(t)$ is a flat connection. 

\section{Conclusion}

We have suggested the way to compare the Fukaya category with the category
of holonomic modules over the quantized algebra of smooth functions.
The idea is for a pair of Lagrangian submanifolds $L_1,L_2$ to find
a kernel $K(L_1,L_2)$ which transforms local systems (or more general
$A_X$-modules) supported on $L_2$ to local systems supported on $L_1$.
We have conjectured that it is possible to make these choices in such a way that
the counting of instantons (i.e. higher compositions in the Fukaya category)
can be replaced by pure algebraic operation of taking homomorphisms
between local systems having the same Lagrangian support. We have checked
the conjecture in the simplest case.
It would be interesting to check it in other cases as well as  ``globalize'' this picture, finding kernels
$K(L_1,L_2)$ for Lagrangian submanifolds which are not close to each other.

\vspace{10mm}

{\bf References}

\vspace{3mm}

[AP] D. Arinkin, A. Polishchuk, Fukaya category and Fourier transform,
math.AG/9811023.

\vspace{2mm}

[Be] V. Berkovich, Spectral theory and analytic geometry over
non-Archimedean fields. Mathematical Surveys and Monographs, 33.
Amer. Math. Soc., 1990.

\vspace{2mm}

[BD] A. Beilinson, V. Drinfeld, Chiral algebras, preprint.

\vspace{2mm}

[BK] S. Barannikov, M. Kontsevich,
Frobenius Manifolds and Formality of Lie Algebras of Polyvector Fields, 
alg-geom/9710032. 

\vspace{2mm}

[De] P. Deligne, Defortmations de l'algebre des fonctions d'une
variete symplectique. Selecta Math. 1 (1995), 667-698.

\vspace{2mm}

[FOOO] K. Fukaya, Y.G.Oh, H. Ohta, K.Ono, Lagrangian intersection
Floer theory-anomaly and obstruction. Preprint Kyoto University (2000).
 
 \vspace{2mm}

[Fu1] K. Fukaya, Floer homology of Lagrangian foliation and 
non-commutative mirror symmetry. Preprint 98-08, Kyoto University, 1998.

\vspace{2mm}

[Gi] V. Ginzburg, Characteristic varieties and vanishing cycles,
Inv. Math. 84 (1986), 327-402.

\vspace{2mm}

 [GLO] V. Golyshev, V. Lunts, D. Orlov, Mirror symmetry for abelian varieties,
 math.AG/9812003.
   
\vspace{2mm} 

[GM] M. Grinberg, R. Macpherson, Euler characteristics and Lagrangian
intersections, in Symplectic Geometry and Topology (ed. Ya. Eliashberg),
1999, 265-294. 

\vspace{2mm}

[K]  A. Kapustin, D-branes in a topologically nontrivial B-field,
  hep-th/9909089 

\vspace{2mm}

[KaMa] M. Karasev, V. Maslov, Pseudo-differential operators and canonical operator
on general symplectic manifolds. Izvestia AN SSSR, 47:5 (1983), 999-1029 (in Russian).

\vspace{2mm}

[KaV], M. Kapranov, E. Vasserot, Vertex algebras and the formal loop space,
 math.AG/0107143.
 
 \vspace{2mm}

[KO] A. Kapustin, D. Orlov, Remarks on A-branes, mirror symmetry and
the Fukaya category, hep-th/0109098

\vspace{2mm}

[Ko1] M. Kontsevich, Homological algebra of Mirror symmetry. Proceedings
of the ICM in Zurich, vol. 1, p. 120-139.

\vspace{2mm}

[Ko2] M. Kontsevich, Deformation quantization of Poisson manifolds, I, 
q-alg/9709040.

\vspace{2mm}

[Ko3] M. Kontsevich, Operads and Motives in Deformation Quantization,
 math.QA/9904055.
 
 \vspace{2mm}

[KR] M. Kontsevich, A. Rosenberg, Non-commutative smooth spaces, 
math.AG/9812158.

\vspace{2mm}

[KoSo1]  M. Kontsevich, Y. Soibelman,  
Homological mirror symmetry and torus fibrations,
math.SG/0011041

\vspace{2mm}

[KoSo2]  M. Kontsevich, Y. Soibelman, Deformation Theory (book, to appear) 

\vspace{2mm}

[KoSo3]  M. Kontsevich, Y. Soibelman, Deformations of algebras over
operads and Deligne's conjecture'', math.QA/0001151.

\vspace{2mm}

[MSV] F. Malikov, V. Schechtman, A. Vaintrob, Chiral de Rham complex,
math.AG/9803041.

\vspace{2mm}

[Se1] P. Seidel, Vanishing cycles and mutations, math.SG/0007115.

\vspace{2mm}

[So1] Y. Soibelman, Quantum tori, mirror symmetry and deformation
theory, math.QA/0011162 and Lett. Math.Phys. v.56 (2001).

\vspace{3mm}

Addresses: 

P.B.:  Department of Mathematics, University of Arizona, 

Tucson, AZ 85721

{email: bressler@hedgehog.math.arizona.edu}

\vspace{2mm}

Y.S.: Department of Mathematics, KSU, Manhattan, KS 66506, USA

{email: soibel@math.ksu.edu}

\end{document}